\documentclass[%
 reprint,
 amsmath,amssymb,
 aps,
]{revtex4-2}

\usepackage{graphicx}% Include figure files
\usepackage{dcolumn}% Align table columns on decimal point
\usepackage{bm}% bold math
\usepackage{hyperref}% add hypertext capabilities
\usepackage{xcolor}
\begin{document}

\preprint{APS/123-QED}

\title{Broad p-Wave Sommerfeld Resonances\\From Dimensional Deconstruction}% Force line breaks with \\

\author{Nobuki Yoshimatsu}
\affiliation{Supreme School for Advanced Education\\1-5-4\ Nishi Takamatsu, Wakayama-shi, Wakayama 641-0051, Japan}
\affiliation{Chiben Gakuen Wakayama Junior/Senior High School \\ 2066-1\ Fuyuno, Wakayama-shi, Wakayama 640-0392, Japan}

% E-mail addresses: only for the corresponding a
\email{nyoshimatsu260@gmail.com}

\date{\today}% It is always \today, today,

\begin{abstract}
We investigate whether dimensional deconstruction can qualitatively modify long-range forces. We show that the deconstruction of a five-dimensional AdS spacetime effectively generates an inverse-square potential, which naturally gives rise to broad p-wave Sommerfeld resonances without fine-tuning. We then apply this mechanism to dark matter phenomenology.
\end{abstract}

\maketitle
\section{Introduction}
Dimensional deconstruction was originally developed as a framework to address the ultraviolet behavior of four-dimensional field theories by interpreting a product gauge group as a discretized extra dimension and providing a four-dimensional description of higher-dimensional gauge theories \cite{ArkaniHamed:2001ca, ArkaniHamed:2001nc, Randall:2002qr}. Later, Son and collaborators \cite{Son2002} showed that the limit of an infinite number of gauge groups reproduces a five-dimensional theory in an AdS spacetime. From a modern perspective, dimensional deconstruction is viewed not merely as a discretization of an extra dimension, but as a four-dimensional realization of higher-dimensional gauge theories and holographic descriptions.
In this study, we adopt this approach to investigate how a product gauge group modifies the behavior of long-range forces. Specifically, assuming that the dark sector possesses product gauge dynamics, we show that in the dimensional deconstruction of a 5D AdS theory, a large number of massive gauge bosons generate an inverse-square potential at short to intermediate distances, which is subsequently experienced by dark fermions, eventually identified with the dark matter (DM). We then demonstrate that this modified scaling of the potential drastically changes the nature of the Sommerfeld effect \cite{Sommerfeld:1931,Hisano:2004ds,Cassel:2009wt}. In particular, the inverse-square potential effectively reduces the centrifugal barrier in p-wave scattering. 
Consequently, we find that Sommerfeld resonances do not require fine-tuning of parameters; rather, they emerge as a natural and generic phenomenon. We also note that the inverse-square potential is a generic consequence of the continuum limit of dimensional deconstruction, while the AdS realization naturally explains the infrared confinement scale.
We then apply this argument to DM annihilation. Regarding, it was pointed out in \cite{Totani:2025fxx} that the gamma-ray spectral signal from the Milky Way halo is in tension with constraints from dwarf spheroidal galaxies if DM preferentially annihilates into $b\bar{b}$. We argue that the Two-Higgs-Doublet Model (2HDM) \cite{Branco:2011iw, Glashow:1976nt} may provide the preferential annihilation into $b\bar{b}$, thus showing that this tension can be resolved by the p-wave Sommerfeld enhancement. Finally, we briefly demonstrate that the DM has nonthermal relics through the late-time production, avoiding its re-annihilation.

\section{Deconstructed Gauge Theory}

\noindent
We consider a product gauge group
\begin{equation}
G
=
SU(M)_1
\times
SU(M)_2
\times
\cdots
\times
SU(M)_N ,
\end{equation}
where adjacent gauge groups are connected by bifundamental link fields
$\Sigma_i$ ($i=1,\cdots,N-1$). The link fields develop vacuum expectation values,
so that the product gauge symmetry is spontaneously broken to the diagonal subgroup,
\begin{equation}
SU(M)_1
\times
SU(M)_2
\times
\cdots
\times
SU(M)_N
\longrightarrow
SU(M)_{\rm diag}\ .
\end{equation}
The gauge bosons acquire a tower of massive states.
For the 5-dimensional AdS extra dimension, the Kalza-Klein (KK) masses are given by \cite{Son2002}
\begin{equation}
    m^2_n=n(n+1)\Lambda^2.
\end{equation}
%%%%%%%%%%%%%%%%%%%%%%%%%%%%%%%%%%%%%%%%%%%%%%%%%%%%
% Gauge interaction of UV-localized dark matter
%%%%%%%%%%%%%%%%%%%%%%%%%%%%%%%%%%%%%%%%%%%%%%%%%%%%
We assume that the DM fermion $\chi$ is localized on the UV site
of the deconstructed gauge theory. Therefore, it transforms only
under the UV gauge group $G_1$, and its interaction is given by
\begin{equation}
\mathcal{L}_{\chi}
=
\bar{\chi}
\left(
i\gamma^\mu D_\mu-m_\chi
\right)\chi ,
\end{equation}
where
\begin{equation}
D_\mu
=
\partial_\mu
-i g_1 A_{1\mu}^a T^a .
\end{equation}
After diagonalizing the gauge-boson mass matrix, the gauge field
at the UV site can be expanded as
\begin{equation}
A_{1\mu}^a
=
\sum_{n=0}^{N-1}
f_n(1) A_\mu^{a(n)},
\end{equation}
where $f_n(1)$ is the corresponding KK wave function evaluated
at the UV site. (Here, the argument \(1\) of \(f_n\) labels the UV site.)
The interaction is reduced to the form of
\begin{equation}
\mathcal{L}_{\rm int}
=
\sum_{n=0}^{N-1}
g_n\,
\bar{\chi}
\gamma^\mu
T^a
\chi\,
A_\mu^{a(n)},
\end{equation}
with
\begin{equation}
g_n
=
g_5 f_n(1),
\end{equation}
where $g_5$ is the five-dimensional gauge coupling.
In the continuum limit, the discrete wave function
$f_n(1)$ is replaced by the bulk wave function evaluated
on the UV brane:
\begin{equation}
g_n
=
g_5\,
f_n(z_{\rm UV}),
\end{equation}
where $f_n(z_{\rm UV})$ denotes the normalized KK wave function
at the UV brane, and 
$f_n(z)$ is expressed as \cite{Son2002}
\begin{equation}
    f_n(z) = -c_n \dfrac{P^1_n(\tanh z)}{\cosh z},\ \ c_n=\sqrt{\dfrac{2n+1}{2n(n+1)}}.
\end{equation}
The exchange of the KK gauge bosons then generates the effective potential
\begin{equation}
V(r)
=
-
\frac{1}{r}
\sum_{n=1}^{N-1}
\dfrac{g_n^2}{4 \pi}
e^{-m_n r}.
\label{eq:KKpotential}
\end{equation}
%For sufficiently short distances satisfying
%$r\ll m_1^{-1}$,
%the KK tower can be approximated by a continuum.
Provided that $z_{UV} \sim 0$, in the intermediate-distance regime $
(Nm_1)^{-1}\lesssim r\lesssim m_1^{-1}$, the potential behaves as
\begin{equation}
V(r)
\simeq
-
\frac{\alpha}{ \Lambda r^2} \cdot \exp(-m_1r),
\label{eq:r2potential}
\end{equation}
where
\begin{equation}
    \alpha=\frac{g_5^2}{4\pi^2 \cosh z_{\rm UV}} \simeq \frac{g_5^2}{4\pi^2} .
\end{equation}
In contrast, at long or extremely short distances, the potential exhibits the usual Yukawa behavior:
\begin{gather}
    V(r) \simeq \begin{cases} -\dfrac{2 \alpha}{ r} \cdot \exp(-m_1 r) \ \ \ \text{for}\ r\gtrsim m_1^{-1}, \notag \\ \notag \\ 
    -\dfrac{N \alpha}{r} \cdot \exp(-m_1 r) \ \ \ \text{for} \ r\lesssim (N m_1)^{-1}. \end{cases}
\end{gather}
Meanwhile, for continuum limit ($N \rightarrow \infty$), we obtain the following behavior:
\begin{equation}
    V_{cont.}(r)
\simeq
-
\frac{\alpha}{\Lambda r^2} \cdot \exp(-m_1r).
\end{equation}
(We discuss the derivation for $V(r)$ in Appendix A.) In FIG.~\ref{fig:potential}, both of the KK sum potential with $N=100$ and continuum potential are displayed.
\begin{figure}[t]
\centering
\includegraphics[width=0.48\textwidth]{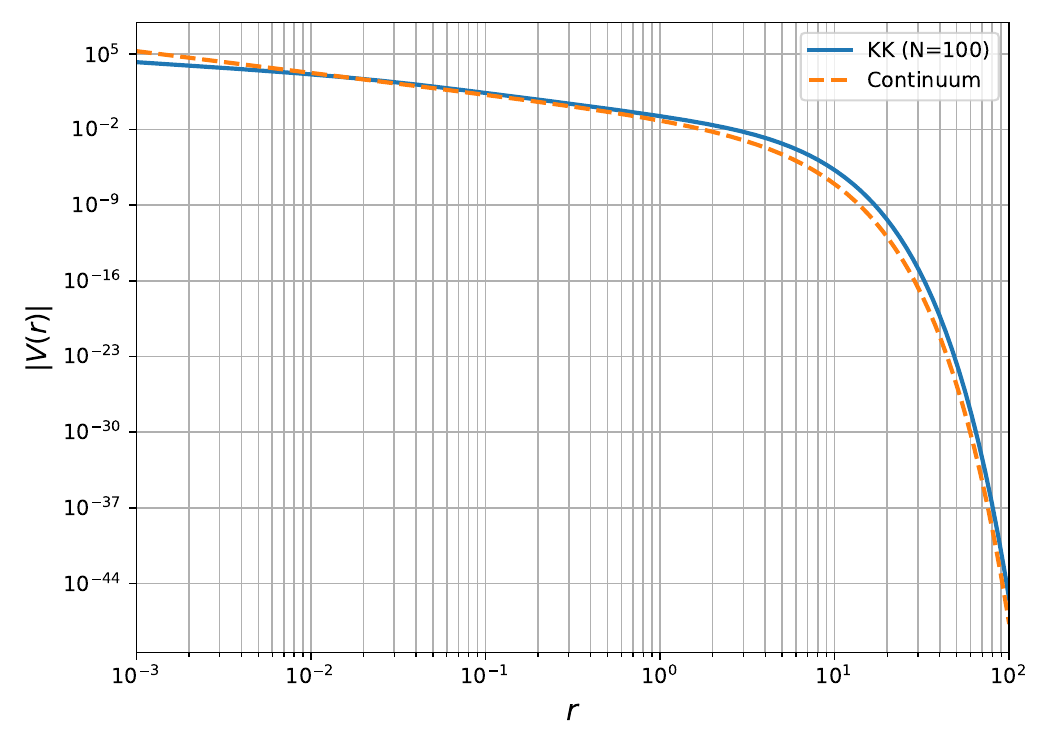}
\caption{
Comparison of the KK sum potential ($N=100$) and continuum potential. We set $\alpha=0.18$ and $\Lambda=0.7$ GeV.
}
\label{fig:potential}
\end{figure}
\section{Generation of Sommerfeld resonance band}

\begin{figure}[t]
\centering
\includegraphics[width=0.48\textwidth]{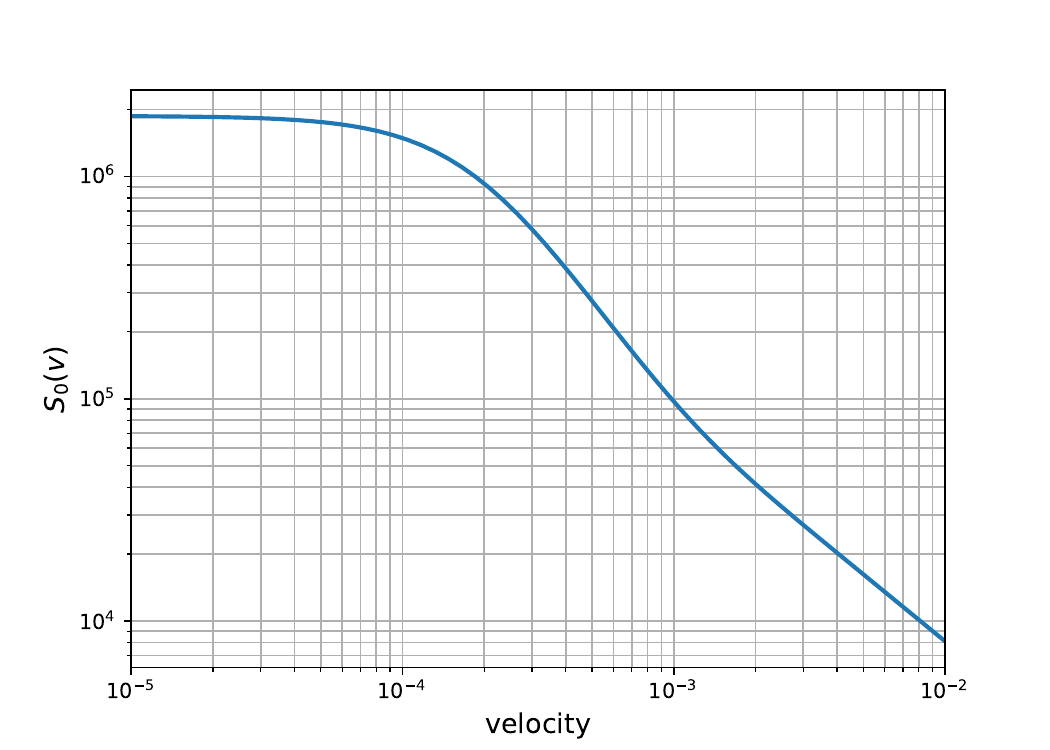}
\caption{
{\it Velocity dependence}:the s-wave Sommerfeld enhancement factor for the KK theory with N=100, setting $\alpha=0.18$ and $\Lambda=0.7$ GeV
}
\label{fig:potential2}
\end{figure}

\begin{figure}[t]
\centering
\includegraphics[width=0.48\textwidth]{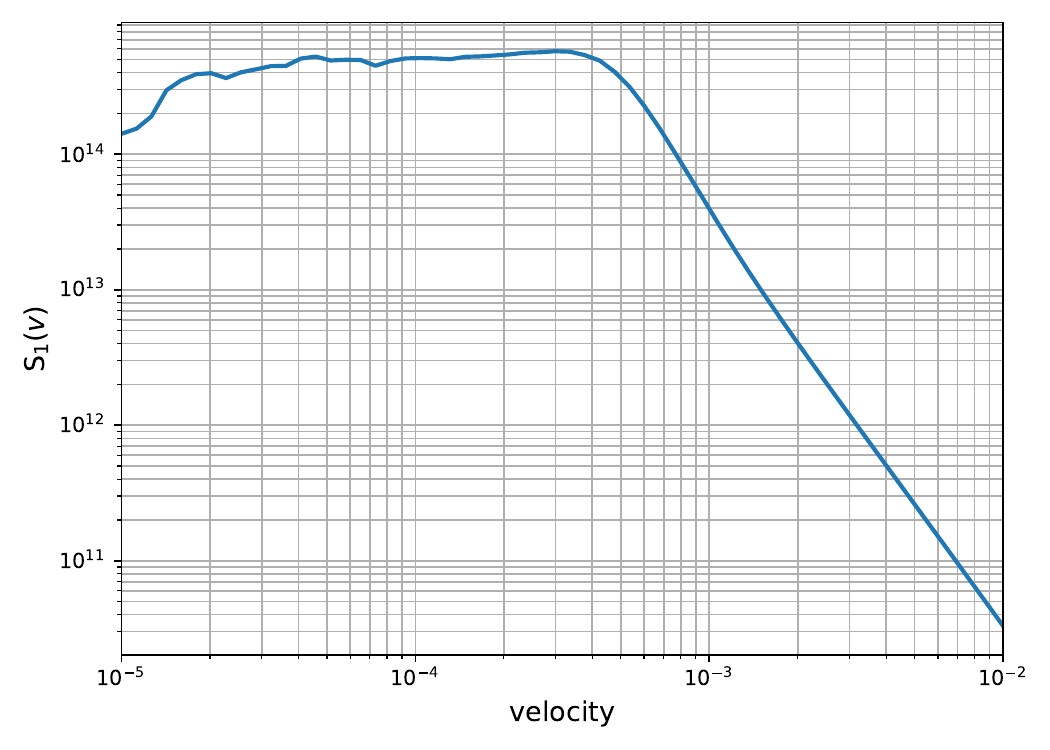}
\caption{
{\it Velocity dependence}: the p-wave Sommerfeld enhancement factor for the KK theory with N=100, setting $\alpha=0.18$ and $\Lambda=0.7$ GeV
}
\label{fig:potential3}
\end{figure}

\begin{figure}[t]
\centering
\includegraphics[width=0.48\textwidth]{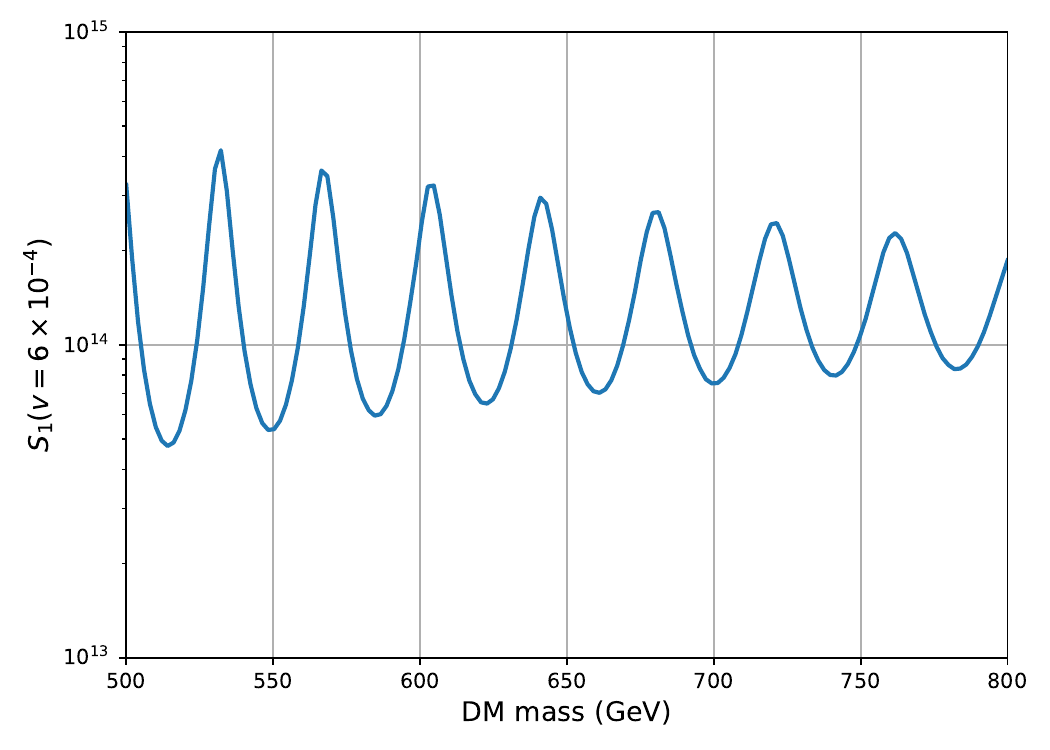}
\caption{
{\it DM mass dependence}: the p-wave Sommerfeld enhancement factor for the KK theory with N=100, setting $\alpha=0.18$ and $\Lambda=0.7$ GeV
}
\label{fig:potential4}
\end{figure}

\noindent
%The radial Schrödinger equation becomes
%\begin{equation}
%u''(r)
%+
%\left[
%k^2
%-
%2\mu V(r)
%-
%\frac{l(l+1)}{r^2}
%\right]
%u(r)
%=0.
%\end{equation}
%Using Eq.~(\ref{eq:r2potential}), one obtains
For simplicity, we use $V_{cont.}$ to analyze the Sommerfeld factor for both s-wave and p-wave. The radial Schrödinger equation is given by
\begin{equation}
u''(r)
+
\left[
k^2
-
\frac{
l(l+1)-2\mu C
}{r^2}
\right]
u(r)
=0,\ \ l=0,1,
\end{equation}
where $k=m_{\chi} v/2$. We stress that for $l=1$ i.e., p-wave,  the inverse-square potential effectively reduces the centrifugal barrier.
The asymptotic radial wave function is written as
\begin{equation}
u_{aym.}(r)
=
A_l\,rj_{l}(kr)
+
B_l\,ry_{l}(kr),
\label{eq:asymptotic}
\end{equation}
where $j_l$ and $y_l$ denote the spherical Bessel and Neumann functions, respectively.
$A_l, B_l$ are obtained analytically as
\begin{align}
A_l
&=
k
\left[
u_{NO}(r_m)
\left(
y_l(x)
+
xy_l'(x)
\right)
-
r_my_l(x)\,
u_{NO}'(r_m)
\right],
\\
B_l
&=
k
\left[
r_mj_l(x)\,
u_{NO}'(r_m)
-
u_{NO}(r_m)
\left(
j_l(x)
+
xj_l'(x)
\right)
\right].
\label{eq:AB}
\end{align}
Here, $x=kr_m$ and $r_m$ is the matching radius, as discussed in Appendix B. $u_{NO}(r)$ expresses $u(r)$ near the origin, which takes the form of
\begin{equation}
    u_{NO}(r) \propto r^{\nu+1}, \ \ \ \nu=\dfrac{-1+\sqrt{(2l+1)^2-8\mu C}}{2}. 
\end{equation}
It is understood that $u_{NO}(r)$ is normalized so that the coefficient of $r^{\nu+1}$ is unity. The Sommerfeld factor $S_l(v)$ is approximately written as follows:
\begin{equation}
    S_{l}(v) \simeq \dfrac{1}{A_l^2+B_l^2}.
\end{equation}
In FIG.~\ref{fig:potential2}, we display the Sommerfeld s-wave enhancement factor for KK theory with $N=100$, while the Sommerfeld p-wave enhancement factor in FIG.~\ref{fig:potential3} and ~\ref{fig:potential4}, both of which show the appearance of the broad resonance band.

%%%%%%%%%%%%%%%%%%%%%%%%%%%%%%%%%%%%%%%%%%%%%%%%%%%%%%%%%%
\section{Effective interaction between dark matter and dark pion}
%%%%%%%%%%%%%%%%%%%%%%%%%%%%%%%%%%%%%%%%%%%%%%%%%%%%%%%%%%
\noindent
Below the confinement scale of $SU(M)_{diag}$, the diagonal gauge interaction
confines the dark quarks into composite states.
The lightest pseudoscalar bound state is identified as the dark pion,
denoted by $\pi_D$.
The interaction between $\chi$ and $\pi_D$
is described by the effective pseudoscalar Yukawa interaction,

\begin{equation}
    \mathcal{L}_{\chi\chi\pi_D}
=
i \dfrac{g_A}{f_{\pi}}
\bar{\chi}
\gamma_5
\chi\partial_{\mu}\pi_D  ,
\label{eq:chidarkpion}
\end{equation}
which is rewritten as
\begin{equation}
\mathcal{L}_{\chi\chi\pi_D}
=
i\dfrac{g_A m_{\chi}}{f_{\pi}}
\pi_D
\bar{\chi}
\gamma_5
\chi ,
\label{eq:chidarkpion}
\end{equation}
Here, $g_A$ denotes the effective coupling generated by the underlying
strong dynamics.
%Since $\pi_D$ is a composite pseudo-Nambu--Goldstone boson,
%the coupling $g_A$ is treated as a phenomenological parameter
%whose magnitude is assumed to be small in our model. 
%%%%%%%%%%%%%%%%%%%%%%%%%%%%%%%%%%%%%%%%%%%%%%%%%%%%%%%%%%
\section{Application to dark matter annihilation}
%%%%%%%%%%%%%%%%%%%%%%%%%%%%%%%%%%%%%%%%%%%%%%%%%%%%%%%%%%
\noindent
Let us apply our argument to the DM annihilation into the standard model particles. 
The interaction between the dark sector and the two Higgs doublets is
assumed to originate from the operator involving the product of link
fields in the moose construction,
\begin{equation}
 {\cal L}_{\rm eff}
 \supset
 c\,\Lambda^2 Tr(\Sigma_1\Sigma_2\cdots\Sigma_{N})
 H_uH_d+{\rm h.c.}
\end{equation}
where the link fields are non-linearly parametrized as 
\begin{equation}
 \Sigma_i=
 \exp\left(\frac{i\pi_i}{f_i}\right).
\end{equation}
The product can be expanded as
\begin{equation}
 \Sigma_1\Sigma_2\cdots\Sigma_N
 \simeq
 \left(
 1
 +i\frac{\sqrt{N}\pi_D}{f_{\pi_D}}
 +\cdots
 \right),
\end{equation}
so that $\pi_D$ is found to 
mix with the $CP$ odd pseudo scalar (denoted $A$):
\begin{equation}
 {\cal L}_{\rm mix}
 \supset
 \mu_{\pi_D A}^{2}\pi_D A,
\end{equation}
where 
\begin{equation}
    \mu_{\pi A}^{2}=\dfrac{c \sqrt{N}\Lambda^2 v_{EW}}{f_{\pi_D}}
\end{equation}
and $v_{EW}=246~{\rm GeV}$ is the electroweak vacuum expectation value, while $f_{\pi{_D}}$ is the dark pion decay constant.
Hence, the corresponding mixing angle is given by
\begin{align}
 \sin\theta_{\pi_D A} &\simeq
 \frac{\mu_{\pi_D A}^{2}}{m_A^2-m_{\pi_{D}}^2} \notag \\& \simeq 2.46\times10^{-2}\,c \cdot \left(\dfrac{N}{100}\right)^{1/2}\left(\dfrac{\Lambda}{0.5\ \text{GeV}}\right)^2 \notag \\ &\ \times \left(\dfrac{0.1\ \text{GeV}}{f_{\pi_D}}\right) \left(\dfrac{0.5\ \text{GeV}}{m_A}\right)^2
\end{align}
for $m_A \gg m_{\pi_{D}}$, where $m_A,\  m_{\pi_D}$ denote each mass of $A$ and $\pi_D$, respectively.

%%%%%%%%%%%%%%%%%%%%%%%%%%%%%%%%%%%%%%%%%%%%%%%%%%%%%%%%%%
\subsection{Dark matter annihilation into $b \bar{b}, t\bar{t}$}
%%%%%%%%%%%%%%%%%%%%%%%%%%%%%%%%%%%%%%%%%%%%%%%%%%%%%%%%%%
\noindent
First, we estimate $\sigma v(\chi\chi
\rightarrow
b\bar b, t\bar{t})$. Dark matter annihilates through the $s$-channel exchange of the dark pion,

\begin{equation}
\chi\chi
\rightarrow
\pi_D^\ast \rightarrow A^\ast
\rightarrow
b\bar b, t\bar{t} .
\end{equation}
The interaction between $A$ and the bottom/top
quark in the type-II two Higgs doublet model \cite{Branco:2011iw} is given by
\begin{equation}
{\cal L}_{Abb}
=
i\frac{m_b}{v}\tan\beta\,
A\,\bar b\gamma_5 b ,\ {\cal L}_{Att}
=
i\frac{m_t}{v}\cot\beta\,
A\,\bar t\gamma_5 t 
\end{equation}
The interaction of $\pi_D$ with each quark is then written as

\begin{equation}
\mathcal{L}_{\pi_D bb}
=
ig_b
\pi_D
\bar{b}
\gamma_5
b,\ \mathcal{L}_{\pi_D tt}
=
ig_t
\pi_D
\bar{t}
\gamma_5
t
\label{eq:darkpionbb}
\end{equation}
where $g_{b, t}$ denotes the effective coupling to the bottom quark, which is given by
\begin{equation}
    g_b= \dfrac{m_b \sin{\theta_\pi}}{v} \tan{\beta},\ g_t= \dfrac{m_t \sin{\theta_\pi}}{v} \cot{\beta}.
\end{equation}
At tree-level, the annihilation cross section of $\chi \chi \rightarrow b\bar{b}$ is given by
\begin{align}
\sigma v(\chi \chi \rightarrow b\bar{b})
\simeq
a + b v^2
\label{eq:sigmavbb}
\end{align}
with 
\begin{align}
a=\frac{N_c}{2\pi}
\frac{
g_A^2 g_b^2
 m_b^2
}
{
\left(4m_\chi^2-m_{\pi_D}^2\right)^2
+
m_{\pi_D}^2\Gamma_{\pi_D}^2
}
\,
\sqrt{1-\frac{m_b^2}{m_\chi^2}}\notag \\
\ \ \ b=\frac{N_c}{8\pi}
\frac{
g_A^2 g_b^2
m_\chi^2 
}
{
\left(4m_\chi^2-m_{\pi_D}^2\right)^2
+
m_{\pi_D}^2\Gamma_{\pi_D}^2
}
\,
\sqrt{1-\frac{m_b^2}{m_\chi^2}},
\end{align}
where $N_c=3$ is the color factor and
$m_{\pi_D}$ is the dark-pion mass.
$\Gamma_{\pi_D}$ denotes its total decay width, and $\left|4m_\chi^2-m_{\pi_D}^2 \right| \gg m_{\pi_D} \Gamma_{\pi_D}$ 
\footnote{The dark pion $\pi_D$ can decay into the Standard Model fermions through its mixing with the pseudo scalar mediator $A$. We parameterize the effective interaction as
\begin{equation}
\mathcal{L}_{\pi_D ff}
\simeq
i\,c_f\frac{m_f}{f_{\pi_D}}
\pi_D\,\bar f\gamma_5 f .
\end{equation}
One derives the corresponding partial decay width as follows:
\begin{equation}
\Gamma(\pi_D\to f\bar f)
=
\frac{N_c}{8\pi}
c_f^2
\frac{m_f^2m_{\pi_D}}{f_{\pi_D}^2}
\sqrt{1-\frac{4m_f^2}{m_{\pi_D}^2}} .
\end{equation}
Given $m_{\pi_D}\sim 0.5~{\rm GeV}$ and
$f_{\pi_D}\sim 0.5~{\rm GeV}$, for example, the
$\mu^+\mu^-$ channel gives the decay width:
\begin{equation}
\Gamma(\pi_D\to\mu^+\mu^-)
\simeq
8.2\times10^{-10}\,\left(\dfrac{c_\mu}{10^{-3}}\right)^2~{\rm GeV}.
\end{equation}
}.
Including the Sommerfeld enhancement,
the annihilation cross section is reduced to the following form:
\begin{equation}
    (\sigma v)_{\rm eff}(\chi \chi \rightarrow b\bar{b})
=
S_0(v)a +S_1(v)\,
bv^2.
\end{equation}
\begin{figure}[t]
\centering
\includegraphics[width=0.48\textwidth]{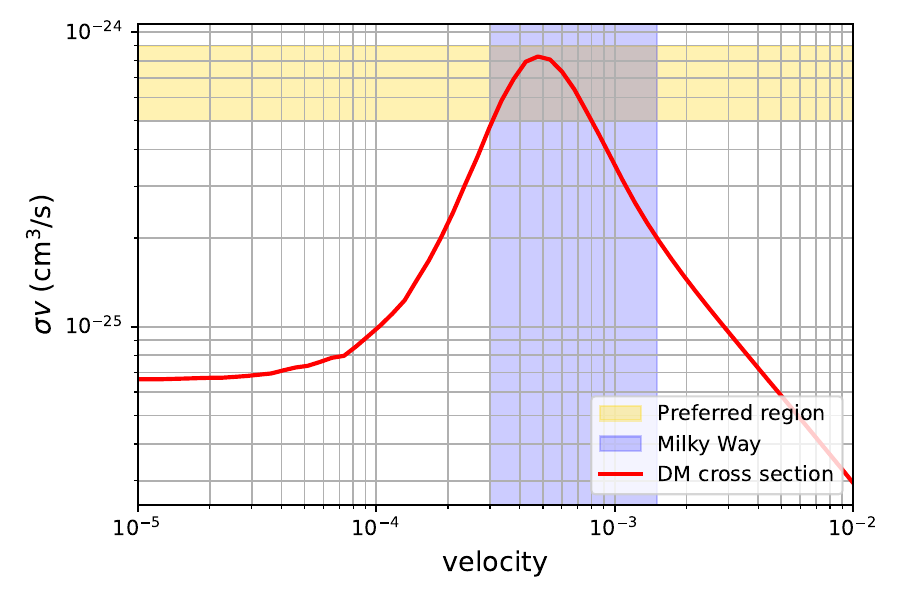}
\caption{
{\it Velocity dependence:} Total cross section of $\chi \chi \rightarrow \bar{b}b$ for $m_{\chi}=600$ GeV. The cross section remains below the current constraints from dwarf spheroidal galaxies at the corresponding low velocities \cite{McDaniel:2023fwk}.
}
\label{fig:potential5}
\end{figure}
\begin{figure}[t]
\centering
\includegraphics[width=0.48\textwidth]{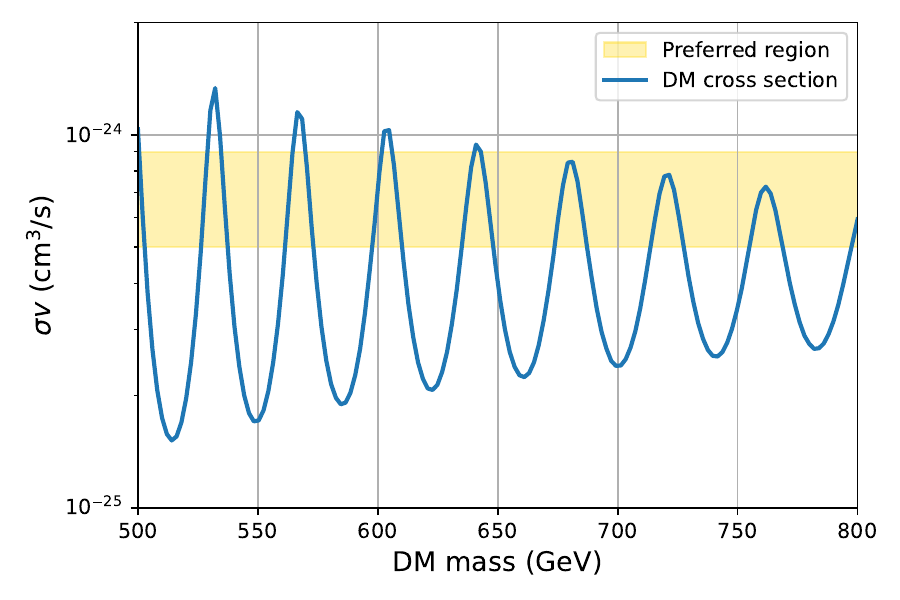}
\caption{
{\it DM mass dependence:} Total cross section of $\chi \chi \rightarrow b\bar{b}$ for $v=6 \times 10^{-4}$. 
}
\label{fig:potential6}
\end{figure}
Meanwhile, the total cross section of $\chi \chi \rightarrow t\bar{t}$ is given by
\begin{align}
  &  (\sigma v)_{\rm eff}(\chi \chi \rightarrow t\bar{t})=\dfrac{m^2_t}{m^2_b \tan^4\beta} \cdot (\sigma v)_{\rm eff}(\chi \chi \rightarrow b\bar{b}) \notag \\
    &=3.5 \times 10^{-2} \left(\dfrac{15}{\tan\beta}\right)^4 \cdot (\sigma v)_{\rm eff}(\chi \chi \rightarrow b\bar{b}).
\end{align}
We thus conclude that for large $\tan \beta$, the $b\bar{b}$ channel dominates the $t\bar{t}$ channel.

%%%%%%%%%%%%%%%%%%%%%%%%%%%%%%%%%%%%%%%%%%%%%%%%%%%%%%%%%%
\subsection{Dark matter annihilation into $W^+W^-, Z^0Z^0$}
%%%%%%%%%%%%%%%%%%%%%%%%%%%%%%%%%%%%%%%%%%%%%%%%%%%%%%%%%%
\noindent
The annihilation channel into electroweak gauge bosons proceeds through
the loop-induced couplings of the pseudo scalar mediator:
\begin{equation}
\chi\chi
\rightarrow
\pi_D^*
\rightarrow
A^*
\rightarrow
WW,\,ZZ .
\end{equation}
In a CP-conserving two-Higgs-doublet model, the $AWW$ and $AZZ$
vertices arise only at one loop. Thus, the effective coupling is
schematically given by
\begin{equation}
g_{AVV}\sim
\frac{g^2}{16\pi^2}\,F_{AVV},
\qquad V=W,Z ,
\end{equation}
leading to
\begin{equation}
(\sigma v)_{\rm eff}(\chi\chi\rightarrow VV)
\propto
|g_{AVV}|^2
\sim
\left(\frac{g^2}{16\pi^2}\right)^2 .
\end{equation}
For large $\tan\beta$, the coupling of $A$ to down-type quarks is
enhanced, while the $AVV$ channels remain loop suppressed.
Consequently, the $b \bar{b}$ channel dominates the $WW, ZZ$ channel:
\begin{equation}
(\sigma v)_{\rm eff}(\chi\chi\rightarrow WW,ZZ)
\ll
(\sigma v)_{\rm eff}(\chi\chi\rightarrow b\bar b).
\end{equation}
In FIG.~\ref{fig:potential5} and ~\ref{fig:potential6}, we display the total annihilation cross section of $\chi\chi \rightarrow \bar{b}b$. Here, we set $\alpha=0.18$, $\Lambda=0.7$ GeV, $\tan \beta=15, \sin \theta_{\pi_D A}=5.0\times 10^{-3}$ and $g_A/f_{\pi_D}=2.5\times 10^{-4}$, while $m_{\chi}=600$ GeV in FIG.~\ref{fig:potential5}, and $v=6.0\times 10^{-4}$ in FIG.~\ref{fig:potential6}.

%%%%%%%%%%%%%%%%%%%%%%%%%%%%%%%%%%%%%%%%%%%%%%%%%%%%%%%%%%
\section{Relic abundance of $\chi$ \\ from late time production}
%%%%%%%%%%%%%%%%%%%%%%%%%%%%%%%%%%%%%%%%%%%%%%%%%%%%%%%%%%
\noindent
The relic abundance of $\chi$ need not originate from thermal freeze-out. As an alternative possibility, $\chi$ can be produced nonthermally through the late-time decay of a slightly heavier long-lived particle,
\begin{equation}
X\rightarrow\chi+\cdots .
\end{equation}
%For instance, a gravitino with a mass of order $1~{\rm TeV}$ can realize such a SuperWIMP-like production mechanism.
To illustrate the condition for the produced $\chi$ population to survive, we consider the annihilation rate:
\begin{equation}
\Gamma_\chi=n_\chi\langle\sigma v\rangle_{\rm eff},
\end{equation}
and compare it with the Hubble expansion rate. During radiation domination, $H(T), \ n_\chi$ are written as
\begin{equation}
H(T)=\sqrt{\frac{\pi^2g_*}{90}}
\frac{T^2}{M_{\rm pl}},
\qquad
M_{\rm pl}=2.4\times10^{18}~{\rm GeV},
\end{equation}
\begin{equation}
n_\chi=Y_\chi s(T),
\qquad
s(T)=\frac{2\pi^2}{45}g_{*s}T^3 .
\end{equation}
Thus, we obtain the following ratio:
\begin{equation}
\frac{\Gamma_\chi}{H}\simeq
1.33 \times
\frac{g_{*s}}{\sqrt{g_*}}
Y_\chi
M_{\rm Pl}T
\langle\sigma v\rangle_{\rm eff}.
\label{eq:Gamma_over_H_late}
\end{equation}
In the present model, the effective annihilation rate can receive contributions from the large number of accessible gauge-boson modes as well as from the Sommerfeld enhancement. Schematically, we may write the annihilation cross section \cite{Dirgantara:2020lqy}:
\begin{equation}
\langle\sigma v\rangle_{\rm eff}
\simeq
N_{\rm mode}
\left[
S_0(v) \frac{28g^4}{432 \pi m_\chi^2}+S_1(v)v^2 \frac{4g^4}{ 576 \pi m_\chi^2}
\right],
\end{equation}
where $N_{\rm mode}$ denotes the effective number of accessible final-state channels. For $N=100$, the number of two-mode combinations is of order:
\begin{equation}
N_{\rm mode}\sim \dfrac{1}{2} \cdot \dfrac{N}{2} \cdot\dfrac{N+1}{2}\simeq 1.3\times10^3 .
\end{equation}
One finds that for $g=1.5,\  m_{\chi}=600$ GeV, the annihilation rate can remain below the Hubble rate for $T\lesssim \mathcal{O}(1)$ eV:
\begin{equation}
\left.\frac{\Gamma_\chi}{H}\right|_{T\lesssim \mathcal{O}(1)\ eV}\ \lesssim1,
\end{equation}
given that the DM recoil velocity at decay of $X$ satisfies $v\lesssim 10^{-4.5}$, implying $S_0 \sim 10^5,\  S_1 v^2 \lesssim \mathcal{O}(10^{5})$. Hence, we note that the required late-time abundance may be realized
through a SuperWIMP-like decay of a metastable parent particle
$X$ with $m_X\gtrsim m_\chi$ \cite{Feng:2003xh, Feng:2003uy}. However, a detailed analysis of the subsequent dark-sector evolution and possible dark-sector reheating is beyond the scope of this work.
%late-time production provides a viable nonthermal origin for a surviving $\chi$ population, without requiring $\chi$ to account for all of the dark matter or to originate from thermal freeze-out. 

\section{Conclusion}
\noindent
In this letter, we discussed the emergence of the inverse-square potential due to the dimensional deconstruction. We then demonstrated that the broad Sommerfeld (specifically $p$-wave) resonance band can appear, assuming the product gauge group of $SU(3)^N$ with $N=100$ for instance.
Subsequently, we applied this argument to a pair of the DM annihilation into $b \bar{b}$, which is larger than those of into $t\bar{t}, W^+W^-$ and $Z^0Z^0$, providing a possible resolution of the tension between
the Milky Way gamma-ray signal and dwarf-spheroidal constraints. Finally, we proposed a late-time production mechanism for the DM at a few eV of the universe temperature, in which the DM can obtain a viable nonthermal abundance while avoiding significant re-annihilation into the KK gauge bosons.  

\appendix

\section{Generation of deconstructed Potential}

\noindent
The deconstructed potential takes the form of
\begin{equation}
    V=-\dfrac{1}{r} \sum_n \dfrac{g_n^2}{4 \pi} \exp(-m_nr),
\end{equation}
where
\begin{equation}
    m^2_n=n(n+1)\Lambda^2, \ g_n=g_5 f_n(z_{UV})
\end{equation}
with
\begin{equation}
    f_n(z)=-\sqrt{\dfrac{2n+1}{2n(n+1)}} \cdot \dfrac{P^1_n(\tanh(z)}{\cosh(z)}.
\end{equation}
For fixed $x=\cos\theta$ ($0<\theta<\pi$), the associated Legendre function has the asymptotic form for $n\gg1$
\begin{equation}
P_n^{1}(\cos\theta)
\simeq
\sqrt{\frac{2n\sin\theta}{\pi}}\,
\sin\!\left[
\left(n+\frac12\right)\theta-\frac{\pi}{4}
\right].
\end{equation}
Therefore, it follows that
\begin{equation}
\left[
P_n^{\,1}(\tanh z_{\rm UV})
\right]^2
\simeq
\frac{n}{\pi \,{\rm sech}\,z_{\rm UV}}
\left[
1-\sin\!\left((2n+1)\theta_{\rm UV}\right)
\right],
\end{equation}
where
\begin{equation}
\theta_{\rm UV}
=
\arccos(\tanh z_{\rm UV}).
\end{equation}
Then one finds
\begin{equation}
g_n^2
\simeq
\frac{g_5^2}{\pi \cosh z_{\rm UV}}
\left[
1-\sin\!\left((2n+1)\theta_{\rm UV}\right)
\right].
\end{equation}
We assume $\theta_{\rm UV}\sim \pi/2$, i.e. $z_{\rm UV} \sim 0$, which leads to the following result:
\begin{align}
V&=-\frac{\alpha}{r} \sum_{n=1}^N\  \left[1-(-1)^n \right]  \exp(-m_nr) \notag \\ &= 
-\frac{2 \alpha}{r} \sum_{k=1}^{N/2} \exp(-m_{2k} r)
\notag \\ &\simeq -\frac{2\alpha}{r} \cdot\dfrac{\exp(-m_1 r)\left[1-\exp(-N \Lambda r)\right]}{1-\exp(-2 \Lambda r)}, 
\end{align}
where
\begin{equation}
    \alpha=\frac{g_5^2}{4\pi^2 \cosh z_{\rm UV}} \simeq \frac{g_5^2}{4\pi^2}.
\end{equation}
Hence, we obtain the asymptotic behavior as follows:  
\begin{gather}
    V(r) \simeq \begin{cases} -\dfrac{2 \alpha}{r} \cdot \exp(-m_1 r)\ \ \ \text{for}\ r\gtrsim m_1^{-1}, \notag \\ \notag \\ 
   -\dfrac{\alpha}{ \Lambda r^2} \cdot \exp(-m_1 r)\ \ \  \text{for} \ (N m_1)^{-1} \lesssim r \lesssim m_1^{-1}, \notag \\ \notag \\  -\dfrac{N \alpha}{ r} \cdot \exp(-m_1 r)\ \ \ \text{for} \ r\lesssim (N m_1)^{-1}.  \end{cases} \tag{A9}
\end{gather}

\noindent
Meanwhile, the potential in continuum limit $(N \rightarrow \infty)$ is given by
\begin{align}
    V=&-\dfrac{1}{r} \int_{m_1}^\infty dm \rho(m) \alpha(m) \exp(-mr) \notag \\ 
    &\simeq -\dfrac{\alpha}{\Lambda r^2} \cdot \exp(-m_1 r), \tag{A10}
\end{align}
where 
\begin{equation}
    \rho(m) \simeq \dfrac{dn}{dm} \simeq \dfrac{1}{2 \Lambda},\ \ \  \alpha(m) \simeq \dfrac{g^2_5}{2 \pi^2}. \tag{A11}
\end{equation}

\flushleft{\ }

\section{Derivation of Sommerfeld enhancement factor}

\noindent
We roughly derive the Sommerfeld enhancement factor. The coefficients $A$ and $B$ are determined by matching the numerical solution
to Eq.~(\ref{eq:asymptotic}) at a sufficiently large radius
$r=r_m$.
Defining
$x \equiv kr_m$,
we write the matching conditions as follows:
\begin{align}
u(r_m)
&=
A\,r_mj_1(x)
+
B\,r_my_1(x),
\\
u'(r_m)
&=
A\!\left[
j_1(x)
+
xj_1'(x)
\right]
+
B\!\left[
y_1(x)
+
xy_1'(x)
\right].
\end{align}
Using the Wronskian identity
\begin{equation}
j_1(x)y_1'(x)-j_1'(x)y_1(x)
=
\frac1{x^2},
\end{equation}
we obtain the coefficients analytically as
\begin{align}
A
&=
k
\left[
u(r_m)
\left(
y_1(x)
+
xy_1'(x)
\right)
-
r_my_1(x)\,
u'(r_m)
\right],
\\
B
&=
k
\left[
r_mj_1(x)\,
u'(r_m)
-
u(r_m)
\left(
j_1(x)
+
xj_1'(x)
\right)
\right].
\label{eq:AB}
\end{align}
In the present model, the numerical solution $u(r)$ is obtained by solving the Schrödinger equation with the KK potential:
\begin{equation}
V(r)
=
- \dfrac{1}{r}
\sum_{n=1}^{N}
\frac{g_n^2}{4 \pi} \cdot e^{-m_nr},
\end{equation}
so that the coefficients $A$ and $B$ contain the collective effects of the entire KK tower.

\bibliographystyle{apsrev4-2}
\bibliography{reference}

\end{document}